\documentclass [a4paper,11pt]{article}
\usepackage[english]{babel}
\usepackage{graphicx,amsmath,amsfonts,amssymb,slashed,upgreek,color,caption,amscd,cite,cancel}

\newcommand{\be}{\begin{equation}}
\newcommand{\ee}{\end{equation}}
\newcommand{\beq}{\begin{equation}}
\newcommand{\eeq}{\end{equation}}
\newcommand{\bea}{\begin{eqnarray}}
\newcommand{\eea}{\end{eqnarray}}

\newcommand{\R}{{}^{(3)}\!R}

\newcommand{\nd}{{\dot n}}

\newcommand{\F}{{\cal F}}

\usepackage[stable]{footmisc}
\def\d{\delta}

\def\MM{M_{*}}
\def\R{{}^{(3)}\!R}

\begin{document}

\begin{center}
\Large{\textbf{Effective Field Theory of Cosmological Perturbations}} \\[1cm] 
 
\large{Federico Piazza$^{\rm a,b}$ and Filippo Vernizzi$^{\rm c}$}
\\[0.5cm]

\small{
\textit{$^{\rm a}$  APC, (CNRS-Universit\'e Paris 7), 10 rue Alice Domon et L\'eonie Duquet, 75205 Paris, France 
}}\\[1mm]

\small{
\textit{$^{\rm b}$ PCCP, 10 rue Alice Domon et L\'eonie Duquet, 75205 Paris, France 
}}\\[1mm]

\small{
\textit{$^{\rm c}$ CEA, IPhT, 91191 Gif-sur-Yvette c\'edex, France \\CNRS, URA-2306, 91191 Gif-sur-Yvette c\'edex, France}}

\vspace{.2cm}

\end{center}

\vspace{2cm}

\begin{abstract}
The effective field theory of cosmological perturbations stems from considering a cosmological background solution as a state displaying spontaneous breaking of time translations and (adiabatic) perturbations as the related Nambu-Goldstone modes. With this insight, one can  systematically develop a theory for the cosmological perturbations during inflation and, with minor modifications, also describe in full generality the gravitational interactions of dark energy, which are relevant for late-time cosmology. 
The formalism displays a unique set of Lagrangian operators containing an increasing number of cosmological perturbations and derivatives. We give an introductory description of the unitary gauge formalism for theories with broken gauge symmetry---that allows to write down the most general Lagrangian---and of the St\"uckelberg ``trick"---that allows to recover gauge invariance and to make the scalar field explicit. We show how to apply this formalism to gravity and cosmology and we reproduce the detailed analysis of the action in the ADM variables. We also review some  basic applications to inflation and dark energy.

\end{abstract}

\newpage 
\tableofcontents
\vspace{.5cm}

\newpage

\section{Introduction}

Cosmological observations strongly indicate that our Universe has undergone two epochs of accelerating expansion, a primordial one---inflation---and a more recent one that is still ongoing---``dark energy". Exploring new models and mechanisms for these two phenomena has been, during the past decade or two, the aim of a frenetic theoretical activity.
Even though characterized by very different energy scales, inflation and dark energy share obvious similarities.
A universal feature of the proposed models is the presence of (at least) one scalar degree of freedom. In the case of inflation such a presence seems in principle inevitable,  because of the need of a dynamical mechanism to exit the accelerated regime and start the standard big-bang phase of decelerated expansion. For what concerns the present acceleration, while a cosmological constant is still a very bright candidate,  it is worth stressing that every concrete alternative to it  does involve, in a way or another, a new dynamical scalar degree of freedom. 

Models of early and late cosmic acceleration can now explain all present observations by adjusting a limited number of parameters.
Despite such a remarkable success, a deeper understanding of the origin of these phenomena is still lacking. The scalar field $\phi(x)$ that those models invoke has no other reason to exist than the one it has been specifically designed for (producing an acceleration), and it rarely shows links with other aspects of the physical realm that are better known or understood. While hoping for theoretical breakthroughs and in anticipation for the wealth of precious data that the upcoming cosmological probes~\cite{euclid1,euclid2,bigboss} 
will provide, it is worth arming ourselves with a formalism that does not rely on the details of any specific model nor on their supposedly fundamental fields, and that can deal directly with observable quantities. 

To this purpose, the general paradigm of effective field theory (EFT) looks the right tool to use. First, EFT allows to deal directly and efficiently with the  degrees of freedom of a physical system that are relevant  at the energy scales of a given experiment. 
Crucially, such degrees of freedom are not necessarily the ``fundamental" fields of the theory. A celebrated example is QCD, a theory of quarks and gluons that at low energy displays only nucleons and pions: those are the degrees of freedom appearing explicitly in the the non-linear sigma model effective Lagrangian~\cite{weinberg}. Moreover, in the EFT paradigm all possible theories compatible with some given symmetry are systematically classified. This makes their effects at low energy transparent and,  at the same time, efficiently parameterizes our ignorance about new physics. 

In this paper we review a powerful EFT formalism that, as a by-product, directly addresses the above mentioned omni-presence of scalar fields in models of cosmic acceleration. Three of its main features/advantages are the following.

\begin{quote}
{\bf 1.} \emph{Cosmological Perturbations as the relevant d.o.f.} : \ In cosmology,  the relevant low-energy degrees of freedom are arguably the \emph{cosmological perturbations} around the homogeneous Friedmann-Robertson-Walker (FRW) background. Among other aspects, cosmological perturbations are responsible for the anisotropies  of the Cosmic Microwave Background (CMB) and for the large scale structures (LSS) that they have later evolved into. According to the standard paradigm, the origin of such fluctuations is seeded in primordial inflation, while their recent evolution is sensitive to the background behavior and possible dynamical features of dark energy. 
\end{quote}

Now that we have identified the relevant degrees of freedom that we want to treat, it remains to understand \emph{how} to do that. Naively, it would look impossible to write a Lagrangian for the perturbations (e.g.~$\delta \phi(x)$) without having solved the background equations first (e.g.~for $\phi_0(t))$. 
Indeed, the task can be naturally addressed if we choose the scalar itself as the time coordinate. This is the so-called ``unitary gauge". In this case the scalar field dynamics gets ``eaten" by the metric and the problem reduces to that of writing the most general Lagrangian for the metric field alone that is compatible with the residual unbroken three-dimensional diffeomorphism invariance. In Sec.~\ref{sec_2} we introduce the unitary gauge in a completely intrinsic way, i.e., without mentioning a scalar field to begin with. The ``top-down" approach---i.e.~starting with a covariant theory, fixing the time coordinate and go to unitary gauge---is discussed in Sec.~\ref{sec_4}.

What makes cosmological perturbations attractive for an effective treatment is also that they are created small and they remain small on the largest cosmological scales. 

\begin{quote}
{\bf 2.} \emph{Expansion in number of perturbations}\, : \  An expansion of the Lagrangian in number of perturbations---rather than in number of supposedly ``fundamental" fields--- is particularly  useful. The lowest order statistics of CMB anisotropies (two- and  three-point functions) can be traced back to a limited number of effective inflationary operators---the effect of the higher-order operators being suppressed by powers of the power-spectrum. As for structure formation, there is an entire range of wavelengths, from the Hubble length $\sim 10^3$ Mpc down to within the non-linear scales of the matter density contrast $\delta \rho_m/\rho_m$, $\sim 10$ Mpc, where the perturbations in the scalar-metric sector are small.  Barring subtle ``screening" effects, those regimes are well described by linear equations and therefore by the quadratic operators in the dark energy Lagrangian expanded in number of perturbations. Higher order operators start becoming important as we move deeper inside non-linear scales.    
\end{quote}

Finally, for both inflation and dark energy, operators with a different number of derivatives are also effective at different scales.

\begin{quote}
{\bf 3.}  \emph{Expansion in number of derivatives}\, : \ A further hierarchy of scales is given by the number of derivatives contained in each operator, higher derivatives being effective at shorter wavelengths. 
\end{quote}

The latter point should be taken with due care. By ``number of derivatives" here we intend the one present in the  \emph{Lagrangian of the true propagating degree of freedom}, obtained after solving the Hamiltonian and momentum constraints. As we show in detail in Sec.~\ref{sec_3}, this is not necessarily the number of derivatives naively appearing in the unitary gauge operators. 

The ``EFT of cosmological perturbations" features a Lagrangian that  has all the good qualities above stated. This formalism made its first appearance in Ref.~\cite{ghost}, where it was used to study the coupling of the ghost condensate to gravity. It was applied to inflation in Ref.~\cite{EFT1,ArkaniHamed:2007ky} and then more systematically developed by  Cheung {\it et al.}~in Ref.~\cite{EFT2}. Other applications of this formalism to inflation are contained in Refs.~\cite{Bartolo:2010bj,Senatore:2010jy,Bartolo:2010di,multifield,Burrage:2010cu,Bartolo:2010im,Creminelli:2010qf,Baumann:2011dt,daniel,Baumann:2011ws,dissipative,Gwyn:2012mw,Green:2013rd,Renaux-Petel:2013ppa}\footnote{A different approach where EFT is successfully applied within the context of cosmological perturbation theory is the so called effective field theory of the large scale structures. It applies to the growth of structures in the Newtonian regime that can be studied by the use of standard (Euclidean) perturbation theory and consists in integrating out the short-scale modes and incorporating their effect on the large-scale dynamics \cite{Baumann:2010tm,Carrasco:2012cv,Pajer:2013jj,Carrasco:2013sva}. See also \cite{Pietroni:2011iz}.}
The extension  of this approach to late-time acceleration was developed in~\cite{Creminelli:2008wc} for a minimally coupled field and in~\cite{GPV,BFPW,Gleyzes:2013ooa,Bloomfield:2013efa} for more general couplings. 

Here is the basic structure of the action:

\be
\begin{split}
\label{total_action}
S \ =\  & \ S_m[g_{\mu \nu}, \Psi_i] \ +\  \int \! d^4x \, \sqrt{-g} \left[\, \frac{\MM^2}{2} f(t) R - \Lambda(t) - c(t) g^{00}   \right. \\[1.2mm]
& + \, \frac{M_2^4(t)}{2} (\delta g^{00})^2\, -\, \frac{m_3^3(t)}{2} \, \delta K \delta g^{00} 
   - \,  m_4^2(t)\left(\delta K^2 - \d K^\mu_{ \ \nu} \, \d K^\nu_{ \ \mu} \right) \, +\, \frac{\tilde m_4^2(t)}{2} \, \R \, \delta g^{00}   \;, \\
     & - \,  \bar m_4^2(t)\, \delta K^2 \, +\, \frac{\bar m_5(t)}{2} \, \R \, \delta K \, + \frac{\bar \lambda(t)}{2}\,  \R^2 + \ \dots \\[1.2mm]
& \left.  + \, \frac{M_3^4(t)}{3 !} (\delta g^{00})^3 \ -\  \frac{\bar m_2^3(t)}{2} (\delta g^{00})^2 \delta K   \ + \ \dots  \right] \;,
\end{split}
\ee
where $\MM$ is the ``bare" Planck mass,  $\delta g^{00} \equiv g^{00} + 1$, $\delta K_{\mu \nu}$ is the perturbation of the extrinsic curvature of the $t = const.$ hypersurfaces, $\delta K$ its trace and $\R$ its three-dimensional Ricci scalar. We assume throughout a spatially flat FRW universe for the background metric so that $\R$ is null at zeroth order. 
The use of three-dimensional  quantities in the second line and below allows a separation between higher derivatives in space and time respectively. In particular, the fact that we have not considered derivatives of such quantities in the action automatically prevents the appearance of higher (more than two) {\it time} derivatives in the equations of motion.\footnote{Because of Ostrogradski theorem \cite{Ostrogradski}, higher time derivatives are usually associated to the propagation of more than one DOF and  ghost-like instabilities. However, from an effective field theory point of view the presence of higher time derivatives is not necessarily a problem, until one specifies the scale at which the ghost shows up, i.e.~its mass \cite{Creminelli:2005qk}. If the latter is much higher than the cut-off scale, higher time derivatives can be treated perturbatively, i.e.~evaluating them using the lower order equations of motion \cite{Simon:1990ic}.} We have not written explicitly operators containing derivatives of these objects.\footnote{An example of an operator containing derivatives of $g^{00}$ is $h^{\mu \nu} \partial _\mu g^{00}  \partial _\nu g^{00} $ where $h^{\mu \nu}$ is the induced metric on the $t= const.$ hypersurface. This operator~\cite{GPV} will not be discussed here but it is of relevance for Lorentz violating models of dark energy and inflation such as~\cite{Horava:2009uw,Blas:2009qj,Blas:2010hb,Blas:2011en,Creminelli:2012xb}.}

 A more technical explanation of the meaning and use of such operators is the subject of the following sections.  Here we give a qualitative description of each line and highlight the aspects formerly advocated for an efficient treatment of inflation and dark energy.

$\bullet$  \emph{The first line} contains the only terms that contribute to the background evolution. In the case of inflation, the matter action $S_m$ is absent and the function $f(t)$ can be set to one by a conformal transformation. On the opposite, in the presence of matter there is a ``preferred" physical~\cite{NP} frame in the approximation---assumed throughout in this paper---that the weak equivalence principle is satisfied\footnote{The choice of a universally coupled metric is stable under radiative corrections in the matter sector~\cite{lamnic,picon,NP}, meaning that WEP violations are generally expected to be delivered by Planck suppressed operators.} and all matter fields are minimally coupled to a ``universal" metric $g_{\mu \nu}$. The fact that \emph{only} three operators, $f(t) R$, $c(t) g^{00}$ and $\Lambda(t)$, contribute to the background evolution is a non trivial consequence of the high number of  symmetries of the FRW solution~\cite{EFT2,GPV}. Explicitly, the background equations derived from action~\eqref{total_action} read~\cite{GPV}
\begin{align}
c \ &  =   \   \frac12 (  -  \ddot f  +  H \dot f ) \MM^2  + \frac12 (\rho_{D}+p_{D}) \;, \label{c2}\\
\Lambda \, &  = \ \frac12 (   \ddot f  + 5 H \dot f ) \MM^2 + \frac12 (\rho_{D}- p_{D})     \;, \label{L2}
\end{align}
where $H(t) = \dot a(t)/a(t)$ is the Hubble constant and we have defined $\rho_D$ and $p_D$ through the equations 
\begin{align}
 H^2  \  &= \  \frac1{3 f \MM^2} (\rho_m + \rho_{D}  )  \label{frie1}\; ,\\
\dot H \, \ &=  \ - \frac1{2 f \MM^2} (\rho_m + \rho_{D} +p_m + p_{D}  )  \label{frie2}\;.
\end{align}
In the above, $\rho_m$ and $p_m$ are the density and pressure of matter fields. Since we are working in the Jordan frame, those behaves as usual (e.g.~$\rho_m \propto a^{-3}(t)$ for non-relativistic matter). Note that in the case of a minimally coupled field $f = 1$. Moreover, as mentioned, for inflation $\rho_m = p_m = 0$ and all the above relations considerably simplify. 

$\bullet$  From \emph{the second line} start the terms in the Lagrangian that do not contribute to the background evolution but only to the perturbations. In particular, the second line contains terms that are quadratic in the number of perturbations and give linear perturbation equations with the lowest number of derivatives (two) for the propagating degree of freedom~\cite{Gleyzes:2013ooa}. The most general scalar-tensor theory with at most two derivatives in the equations of motion---also known as Horndeski theory ~\cite{horndeski,fab-four}, or generalized Galileons~\cite{NRT,Deffayet:2009mn,Deffayet:2009wt}---expanded to second order in perturbations is contained in the second line with the further constraint $m_4 = \tilde m_4$ \cite{Gleyzes:2013ooa,Bloomfield:2013efa}. Being able to reproduce the linear dynamics of such complicate Lagrangians with only six operators is another non trivial result and a great advantage of this approach. 

$\bullet$  \emph{The third line} displays also some quadratic operators (same number of perturbations as the second line) but with higher \emph{spatial} derivatives. They modify the dispersion relation of the propagating degree of freedom with terms that become important at high energy---typically, $k^2$ corrections to the linear relation $w = c_s |\vec k|$.

$\bullet$  Finally, \emph{the fourth line} is a sample of possible \emph{cubic} operators i.e.~higher order in the number of perturbations.

\section{Unitary gauge}
\label{sec_2}

In the next subsection we review very concisely the spontaneous symmetry breaking of global and local symmetries in the case of gauge theories. While referring to ~\cite{weinberg,peskin} for exhaustive and more comprehensible treatments of these subjects, here we just go through the basic logical skeleton of ideas.  In due course, we highlight with Latin numbers the five main points that will be useful later on for applications to cosmology (Sec.~\ref{sec:gravity}).

\subsection{Generalities} \label{Sec_2.1}

Anytime we search for stable, universal and model-independent statements in physics, symmetry is a good direction to look to. In particular, the Nambu-Goldstone phenomenon associated with the spontaneous breaking of a continuous global symmetry allows to grasp the low energy spectrum and dynamics of a theory in a completely general way, \emph{i.e.}  without knowing the details at higher energy of the theory itself.  

Consider a multiplet of $N$ fields $\Phi_i(x)$ whose theory (i.e.~the action) is invariant under ``rotations" in some matrix group $G$, 
\begin{equation} \label{linear}
\Phi_i(x) \ \rightarrow \ \gamma_{ij} \Phi_j(x)\, , \qquad \gamma_{ij} \in G\, .
\end{equation}
If the fields acquire vacuum expectation values $\langle \Phi_i\rangle_0 \neq 0$---which we assume constant in space and time---they will in general ``spontaneously break" the symmetry $G$, in the sense that there will be some elements $\gamma_{ij}$ of the group $G$ such that $\gamma_{ij} \langle \Phi_j \rangle_0 \neq 0$. Quite intuitively, field configurations that differ from the vacuum by a symmetry transformation will be distinct ``equivalent vacua", all at the same energy. In order to explore such configurations we can act on $\langle \Phi_i\rangle_0$ with a generic group element that breaks the symmetry. Such a group element can be written as an exponential of a combination of the \emph{broken  generators} $\tau_a$ of the symmetry group, $\gamma_{ij} \ \sim \ (e^{i \pi_a \tau_a})_{ij}$. We now promote the parameters at the exponent to fields, $\pi_a \rightarrow \pi_a(x)$, 
\begin{equation} \label{gammadef}
\gamma (x) \ = \ e^{i \pi_a(x) \tau_a} \, .
\end{equation}
Note that  the $\pi_a$ parameterize a subset of all possible field configurations and that they must correspond to massless excitations, because in the limit of zero gradients they just interpolate between different but energetically equivalent vacua. As opposed to the original fields $\Phi_i$ that transform linearly~\eqref{linear} under the entire symmetry group $G$, the Goldstones $\pi_a$ transform linearly only under the unbroken subgroup of $G$ that leaves the vacuum invariant, and are in general non-linear realizations of $G$ itself~\cite{Coleman:1969sm}. 
Once the field space $\Phi_i$ is parameterized in terms of the Goldstone fields $\pi_a$, other (``radial") directions will generally be heavy and thus decouple from the low-energy theory. There is also the possibility of other ``accidentally" flat directions in field space. The corresponding fields (``moduli"), however, are not protected by any symmetry and will generally acquire a mass by the effect of quantum corrections. We conclude that the entire low-energy dynamics is encoded in the Goldstone fields $\pi_a(x)$.
Up to a limited number of free coefficients, such a dynamics is entirely fixed by the symmetry breaking pattern.

In order to promote a global symmetry to a local (or gauge) symmetry, we need to introduce covariant derivatives, $\partial_\mu \rightarrow \nabla_\mu = \partial_\mu + i g A_\mu$. We impose that the gauge fields $A_\mu$, that can be seen as matrices acting on the original multiplet $\Phi_i$, transform in such a way to counterbalance the effect that such spacetime dependent transformations have on the derivative terms: 
\begin{equation}\label{stuecksu2}
A_\mu(x) \rightarrow \gamma(x) \left(A_\mu(x) - \frac{i}{g} \partial_\mu \right) \gamma(x)^\dagger\, . 
\end{equation}
It is known that such a prescription produces couplings of the type $\sim g^2 (\Phi_i^* \Phi_i) \, A_\mu A^\mu$. When the symmetry is broken, these are effectively mass terms for the gauge fields. 

As in the case of global symmetries, we can still parameterize the $\Phi_i$ sector with the Goldstones $\pi_a$ plus other heavy ``radial" fields. 
Remarkably, the gauge fields $A_\mu(x)$ and the Goldstones $\pi_a(x)$ are \emph{redundant}, in the sense that different configurations of $A_\mu$ and $\pi_a$ related by a gauge transformation correspond to the same physical situation. 
Among the possible gauge fixing conditions to get rid of such a redundancy, 
\begin{quote}
{\bf (I) } \emph{the unitary gauge is defined by simply setting all Goldstone fields to zero. }
\end{quote}
In the well-known example of the electro-weak theory the gauge group is $SU(2)\times U(1)$ spontaneously broken to a diagonal $U(1)$. In this case, the complex Higgs doublet is reduced, by the unitary gauge prescription, to a single real ``radial" component:
\begin{equation} \label{unitarysu2}
\Phi_j(x)  \ =\ \left(\begin{matrix}
\Phi_1(x) \\[2mm]
\Phi_2(x) \\
\end{matrix} \right) \ \ \rightarrow \ \  \left(\begin{matrix}
v + h (x) \\[2mm]
0 \\
\end{matrix} \right)\, .
\end{equation}
In other words, the unitary gauge prescription picks up, among all equivalent configurations, the particular representative of the $\Phi_i$ sector that does not contain any fluctuation along the symmetry direction.  
In Weinberg's words~\cite{weinberg}, this choice ``makes manifest the menu of physical particles in the theory": all the fields have a straightforward particle interpretation, they directly represent physical states with well-defined (positive) probabilities, from which the name ``unitarity" or ``unitary". 
Also interactions are particularly transparent in the unitary gauge, most physical processes appearing already at tree-level as interaction terms in the Lagrangian.

Quite intuitively, since a precise gauge choice has been made,
\begin{quote}
{\bf (II) } 
\emph{a Lagrangian written in unitary gauge is no longer invariant under the broken symmetries, while it is still invariant under the unbroken symmetries.}
\end{quote}
The above can serve as a guidance to parameterize our ignorance about higher energy physics. With the massive vectors $A_\mu^a$ and the various matter fields (leptons and quarks) we can assemble effective Lagrangians to parameterize the physics beyond the standard model.
Such a Lagrangian can be organized in a series of terms of increasing inverse powers of a ``high energy" scale $\Lambda$, typically that of new massive degrees of freedom, or of new physics. Low energy observables (amplitudes, decay rates etc.) can then be systematically calculated as a power series in $E/\Lambda$, $E$ being the energy relevant for the process in question.

 For other uses---such as understanding the behavior of the theory at  high energy---other gauge choices are more convenient.
 \begin{quote}
{\bf (III) } 
\emph{Starting from a Lagrangian written in unitary gauge,  gauge invariance can be restored by the ``St\"uckelberg mechanism", i.e., by forcing on the fields a gauge transformation that reintroduces the Goldstones.}
\end{quote}  In a way, this is like ``undoing" the gauge fixing~\eqref{unitarysu2}. The Goldstones fields $\pi_a$ will reappear by forcing on the vector bosons $A_\mu$  the transformation~\eqref{stuecksu2} with $\gamma$ defined in~\eqref{gammadef}.
For example, consider a Lagrangian for massive vector bosons in unitary gauge,  
\be
\label{L_A}
{\cal L} = - \frac14 \text{Tr} F_{\mu \nu} F^{\mu \nu} - \frac12 m^2 \text{Tr} A_\mu A^\mu \;, 
\ee 
By applying~\eqref{stuecksu2} and expanding at quadratic order in the Goldstones $\pi$, we obtain
\be
\label{L_A_quadratic}
{\cal L} = - \frac14 \text{Tr} F_{\mu \nu} F^{\mu \nu} - \frac12 (\partial_\mu \pi_c )^2- \frac12 m^2 \text{Tr} A_\mu A^\mu + i m \partial_\mu \pi_c A^\mu\; , 
\ee 
where we have defined the canonically normalized  Goldstone fields $\pi_c \equiv (m/g) \pi$.
 By introducing redundant degrees of freedom, one can just \emph{make} gauge invariant a theory of massive vector fields $A_\mu$.

Having emphasized that the gauge redundancy is not really a symmetry, one might then wonder what it really means, after all,  to spontaneously break it. The issue is subtle and beyond the scope of the present review. Naively, one can always keep as a reference the ``global part" of the symmetry and say that 
\begin{quote}
{\bf (IV) } 
\emph{the gauge symmetry is spontaneously broken if its ``global part" is.}
\end{quote}
In the above example, having a vacuum expectation value in the $\Phi$ sector different from zero, $\langle \Phi_i\rangle_0 \neq 0$, is a gauge invariant statement. Whatever field configuration we choose to represent it, such configuration will break, at least formally, the \emph{global} $SU(2)\times U(1)$. 
More pragmatically, one can also look at the action in the unitary gauge and say that 
\begin{quote}
{\bf (V) } 
\emph{the gauge symmetry is ``spontaneously broken" if, by applying the St\"uckelberg ``trick"~\eqref{stuecksu2} to the action written in unitary gauge, interacting Goldstone particles $\pi_a$ are produced.}
\end{quote}

Finally, note that in the limit $m \to 0$ and $g \to 0$ keeping $m/g$ constant, the Goldstone bosons decouple from the gauge fields $A_\mu$. In other words, at high energies $E \gg m$ it is convenient to use $\pi$ to describe the scattering of massive vector fields. In writing \eqref{L_A} as \eqref{L_A_quadratic} we have neglected cubic and higher-order terms in $\pi_c$ suppressed by $m^2/g^2$, suggesting that the Goldstone boson self-interactions become strongly coupled at energies $E \gg 4 \pi m/g$. The decoupling limit is thus well-defined in the regime $m \ll E \ll 4 \pi m /g$.

\subsection{Cosmology as Spontaneous Symmetry Breaking}
\label{sec:gravity}

General Relativity is a gauge theory because of its invariance under coordinate changes, $x^\mu \rightarrow {x'}^\mu  = {x'}^\mu (x^\nu)$---the metric field $g_{\mu \nu}$ playing the role of the gauge fields sector. In a cosmological context it is useful, in particular, to look at time reparameterizations, $t \rightarrow t'  = t' (x^\nu)$. How can we say whether such a symmetry is spontaneously broken? According to point {\bf (IV)} above, we should start by looking at its global version. A global symmetry of Minkowski space is time translations. More generally, any timelike killing vector defines a global symmetry. Therefore, in the sense above specified, we can say that time-translations are broken by any solution---any spacetime---that does not have a time-like killing vector field. Since we are interested in the local dynamics, it is not too important whether this time-like killing vector is defined globally or not. For instance, de Sitter space is obviously \emph{not} a static solution. However, among its many killing vectors, we can choose one that is time-like in a finite patch. In the usual ``cosmological" coordinates 
\begin{equation}
ds^2 = -dt^2 + e^{2 H t} d \vec x^2
\end{equation}
consider the dilation isometry 
\begin{equation} 
t \ \rightarrow \ t + \Delta t\, , \qquad \vec x \ \rightarrow  \ e^{- H \Delta t} \, \vec x\, .
\end{equation} 
The corresponding killing vector is time-like in an entire finite patch around the origin of the coordinates, until we hit the horizon, $|\vec x| e^{H t} < H^{-1}$. Therefore, for all practical purposes, de Sitter space is a state of gravity with \emph{unbroken} time translations.

The example of de Sitter is relevant because it represents the limiting case of most inflationary models. Crucially, the expansion during inflation is quasi-de Sitter but not quite so, because of the empirical evidence of a red tilt in the primordial power spectrum~\cite{planck} and because we need to exit the accelerating phase at some point, and it is quite natural to think this transition to happen smoothly. Since inflation is not completely de Sitter, we deduce that it must be accompanied with a Goldstone excitation. This is because  we have argued above in point {\bf (V)} that any spontaneous breaking of time translations---or of any gauge symmetry for that matter---is associated with a Goldstone excitation $\pi(x)$ upon application of the St\"uckelberg trick. This field must transform linearly under the unbroken space translations and rotations, which simply means, in this case, that it must be a three-dimensional scalar. It is not difficult to show (e.g.~Ref.~\cite{Nicolis:2011pv}, Sec. 3.3) that $\pi(x)$ can always be ``completed" into a proper four dimensional scalar field. Therefore, rather than postulating a scalar field \emph{ab initio}, the EFT of cosmological perturbations shows  that the presence of a scalar is just the inevitable consequence of broken time translations. This is a powerful point of view to address one's possible unease about postulated ``fundamental" scalars in inflationary theories. 

To build the EFT of cosmological perturbations we start working in unitary gauge. The Goldstone field is absent by the definition {\bf (I)} and, in the case of inflation, matter fields are not there either. We deduce that the \emph{minimal} model of inflation is described in unitary gauge by an action for the metric field alone. By point {\bf (II)}, the unitary gauge action must be invariant under the unbroken symmetries of the problem. As we have argued, a FRW background that is neither Minkowski nor de Sitter only breaks time translations and boosts, but leaves spatial diffeomorphisms unbroken. 

The EFT of cosmological perturbations can thus contain~\cite{EFT2}
\begin{enumerate} 
\item four-dimensional diff-invariant scalars (e.g.~any curvature invariant such as the Ricci curvature $R$) in general multiplied by functions of the time $t$. 
\item four-dimensional covariant tensors with free upper $0$ indices such  as $g^{00}$, $R^{00}$ etc. All spatial indices must be contracted.
\item three-dimensional objects belonging to the $t=constant$ surface, such as the extrinsic curvature $K^{ij}$ and its trace $K$, the three dimensional curvatures $\R$, $\R_{ij}$ etc.
\end{enumerate} 
The last two points deserve a bit more of explanation. By breaking the invariance under time-reparametrization we are allowed to write functions of the metric $g_{\mu \nu}$ that contain information about the specific choice of the chosen time coordinate. We can thus contract covariant tensors with the unitary vector orthogonal to the $t = const.$ surfaces,  
\begin{equation} \label{nmu} 
n_\mu \ = \ - \frac{\delta_\mu^0}{\sqrt{-g^{00}}}\; ,
\end{equation}
thereby producing free upper $0$ indices. But we can also use geometric quantities describing such a surface. By defining the induced metric $h_{\mu \nu} = g_{\mu \nu} + n_\mu n_\nu$,
we can use the extrinsic curvature
\begin{equation} \label{definition_ext}
K_{\mu \nu} \equiv  h^{\ \sigma}_\mu \, \nabla_\sigma n_\nu\; ,
\end{equation}
as well as the three-dimensional Ricci tensor $\R_{\mu \nu} [h_{\rho \sigma}]$.

\subsubsection{Inflation}

Before starting to write an action with all possible combinations of the above ingredients, it is worth   trying to address, at the same time, another important point. Now that we are left with the metric field as the only dynamical variable, it is straightforward to write such an action also already expanded in number of perturbations, which is one of the main \emph{desiderata} expressed in the introduction. At zeroth order in perturbations, the metric is just that of a spatially flat FRW solution,
\begin{equation}
ds^2 = -dt^2 + a^2(t) \, d \vec x^{\, 2}\, ,
\end{equation}
and some of the basic ingredients listed above read (in order of appearance)
\begin{equation}
R_{(0)} = 12 H^2 + 6 \dot H,\qquad  g^{00}_{\ \ (0)} = -1,\qquad K^{i j}_{(0)} = H \delta^{ij}, \qquad K_{(0)} = 3 H, \qquad {\rm etc.}  
\end{equation}
We can thus start writing a completely general EFT action for inflation, naively of the form
\begin{equation}
S = \int dt \, d^3x\, \sqrt{h} \left[\Lambda_0(t) + c_1(t) ( g^{00} + 1) + c_2(t) (K - 3H(t))  + c_3(t) (R - R_{(0)}(t)) +  \dots \right]\, ,
\end{equation}
where $\sqrt{h}$ is the three-dimensional volume element that takes care of the invariance under 3-d diffeomorfisms, $\Lambda_0$ is the zeroth order term in the perturbations, and the other terms start at first order in the perturbations. 

However, in practice, it is very convenient to rearrange the terms above in a slightly different way. First, it is convenient to use directly the 4-d volume element $\sqrt{-g} = \sqrt{h}/\sqrt{-g^{00}}$ instead of $\sqrt{h}$. This is useful whenever we need to integrate by parts 4-d covariant derivatives or just derive Einstein equations by variations of the action with respect to $g_{\mu \nu}$. Moreover, $d^4 x \sqrt{-g}$ is the invariant volume element and, as we will see, is left unaffected by the St\"uckelberg trick.  Related to this last point, it is also useful to have the combination $\sqrt{-g} R$ sticking out, and merging $R_{(0)}(t)$ together with the zeroth order piece. Moreover, in the absence of matter fields, it is always possible to absorb the time dependent coefficient $c_3(t)$ by redefining an Einstein metric through a conformal transformation. 
Finally, 
by using~\eqref{definition_ext}, the term linear in the extrinsic curvature $K$ can be integrated by parts giving a function of $g^{00}$,
\be
\int d^4x \sqrt{-g} \, \F(t) K=-\int d^4x \sqrt{-g} \, n^\mu \nabla_\mu {\cal F}(t) =-\int d^4x \sqrt{-g} \sqrt{-g^{00}} \dot {\cal F}(t)\, .
\ee
We are thus lead to the following action 
\begin{equation} \label{example2}
 S =  \int d^4x \sqrt{-g} \left[\frac{M_{\rm Pl}^2}{2} R - \Lambda(t) - c(t) g^{00}\right] + S^{(2)}\; .
\end{equation}
In the above, $S^{(2)}$ is made by terms that start already at quadratic order in the number of perturbations. The fact that only three operators (and only two tunable functions of the time, $c$ and $\Lambda$) determine the background evolution is a non-trivial consequence of the symmetries of FRW. A rigorous derivation of such a result is contained in Appendices A and B of Ref.~\cite{EFT2}.

\subsubsection{Dark Energy}

As already stressed, one of the insights of the EFT of inflation is the inevitability of a propagating scalar degree of freedom on a general FRW background that is not de Sitter or Minkowski---for which time translations are unbroken. Indeed, such inevitable scalar fluctuations are nothing else than the \emph{adiabatic perturbations}, as will be clearer from the discussion of the St\"uckelberg mechanism in Sec.~\ref{stuck}. When extending this formalism to late time cosmology, one has to decide how to involve matter fields (baryons, dark matter, radiation etc.)  in the game. In fact, it is convenient to apply this formalism (and thus to write the most general Lagrangian for the metric $g_{\mu \nu}$ in unitary gauge etc.) to the dark energy-gravitational sector only of the theory. For the matter sector we will assume that the weak equivalence principle (WEP) is valid, so that matter fields $\psi_m$ couple to the metric $g_{\mu \nu}$ universally and through a covariant action $S_m[g_{\mu \nu},\psi_m]$. In other words, we assume the existence of a ``Jordan metric" $g_{\mu \nu}$ and we will work with that. It would be more complicated but technically straightforward to consider different matter sectors coupled to different metrics.

This marks the main difference with the case of inflation: if we want to stick with the Jordan frame metric that minimally couples to matter we now need to allow a general free function of time $f(t)$ in front of the Ricci scalar in equation~\eqref{example2}, 
\begin{equation} \label{example3}
 S \ = \ S_m[g_{\mu \nu}, \Psi_i] \  + \int d^4x \sqrt{-g} \left[\frac{\MM^2}{2} f(t) R - \Lambda(t) - c(t) g^{00}\right] + S_{DE}^{(2)}\; .
\end{equation}

 \subsection{Higher Order terms}
 
 The part of the action that contributes at quadratic and higher order in~\eqref{example2} and~\eqref{example3} can be read off from the second, third and fourth line of~\eqref{total_action}. It contains terms such as $\delta g^{00} = g^{00}+1$, quantities of first order in the perturbations. In this respect, it is useful to \emph{define} the perturbation of the extrinsic curvature as
 \begin{equation}
 \delta K_{\mu \nu} = K_{\mu \nu} -  H h_{\mu \nu}, 
 \end{equation}
 where $h_{\mu \nu} = g_{\mu \nu} + n_\mu n_\nu$ is the (perturbed) three dimensional metric of the $t = const.$ surface. 
 
Apart from the Einstein-Hilbert term in the background part of the action (first line), instead of four-dimensional Riemann, Ricci tensors and their contractions we find it convenient to deal  with three dimensional quantities belonging to the $t = const.$ surface (${}^{(3)}\!R_{\alpha \beta \gamma \delta}$, ${}^{(3)}\!R_{\mu \nu}$ and $K_{\mu \nu}$) because they do not explicitly contain higher time derivatives. 
In order to relate four-dimensional with three-dimensional one can make use of the  Gauss-Codazzi equation~\cite{Poisson,Wald:1984rg},
\be
\label{GC_equ}
{}^{(3)}\!R_{\alpha \beta \gamma \delta} = h^\mu_\alpha h^\nu_\beta h^\rho_\gamma h^\sigma_\delta R_{\mu \nu \rho \sigma} - K_{\alpha \gamma} K_{\beta \delta} + 
K_{\beta \gamma} K_{\alpha \delta} \;,
\ee
and its contracting forms.

 \section{St\"uckelberg mechanism} 
 \label{stuck}

We now discuss how to restore gauge invariance and write the same theory in different gauges, with different choices of the coordinates. According to point {\bf (III)} of Sec.~\ref{Sec_2.1} we have to ``force" the broken gauge transformation on the fields written in unitary gauge. Since we have fixed the time coordinate, we have to impose a time coordinate transformation on our action,
\be
\begin{split}
t \to \tilde t &= t+\pi(x^\mu) \label{coord_t} \;, \\ 
x^i \to \tilde x^i &= x^i \;.
\end{split}
\ee
Under this coordinate change a time dependent function in the action, ($f(t)$, $c(t)$ etc.), transforms as
\be
f(t) \to f(\tilde t) = f(t+\pi(x)) = f(t) + \dot f (t) \, \pi(x)  + \ldots \;,
\ee
or, in short,
\be
f(t) \to  f(t) + \dot f \pi + \ldots \;.
\ee
By definition, a scalar does not transform under change of coordinates, e.g.
\be
R( x^\mu ) \to \tilde R(\tilde x^\mu) = R(x^\mu) \; .
\ee
The same holds true for the volume element $d^4 x \sqrt{-g}$ as well as for the entire matter action if it is covariant and universally coupled to the Jordan metric, as we have assumed\footnote{In general, models written in the Einstein frame or featuring explicit violations of the WEP~\cite{Gasperini:2001pc}, or of Lorentz invariance~\cite{Blas:2012vn} in the dark matter sector represent counterexamples.}.  However, note that for $\delta R \equiv R - 
R^{(0)}( t )$ we have 
\be
\delta R \to  \delta R  - \dot R^{(0)}  \pi + \ldots \;.
\ee
For the contravariant and covariant components of a tensor we have, 
\be
{T}^{\alpha\beta} \to  (\delta^\alpha_\mu + \delta^\alpha_0 \partial_\mu \pi) (\delta^\beta_\nu + \delta^\beta_0 \partial_\nu \pi) T^{\mu \nu}
\;,\label{contrava}
\qquad
T_{\mu \nu} \to   
(\delta^\alpha_\mu -   \delta_0^\alpha \partial_\mu \pi+ \ldots) (\delta^\beta_\nu - \delta^\beta_0  \partial_\nu \pi+ \ldots) T_{\alpha \beta}  \;. 
\ee

In dealing with three-dimensional quantities that are characteristic of the $t= const.$ surface, such as the extrinsic or intrinsic curvatures $K_{\mu \nu} $ and $\R_{\mu \nu}$, it is worth noting that under a change of coordinates they do not just transform covariantly. They truly change as geometrical quantities, because the corresponding surface that they are referring to changes. The spatial components of the extrinsic curvature orthogonal to the constant time hypersurface are given by
\be
K_{ij} = \frac12\sqrt{- g^{00} } (\partial_0 g_{ij} - \partial_i g_{0j}  - \partial_j g_{i0} ) \;. \label{ADM}
\ee
To linear order we can transform each component of the metric in this expression using eq.~\eqref{contrava}, obtaining
\be
\begin{split}
K_{ij} (x^\mu) \to \tilde K_{ij} (\tilde x^\mu)= &\ \frac12\sqrt{- g^{00} } (1+ \dot \pi ) \left[ (1-\dot \pi) \partial_0 g_{ij} - \partial_i (g_{0j}+ \partial_i \pi)  - \partial_j (g_{i0}+ \partial_i \pi) \right] \\
= &\ \frac12\sqrt{- g^{00} } ( \partial_0 g_{ij} - \partial_i g_{0j}  - \partial_j g_{i0})- \partial_i \partial_j \pi \\
=& \ K_{ij} - \partial_i \partial_j \pi\;,
\end{split}
\ee
where $K_{ij}$ in the last line is the extrinsic curvature orthogonal to the constant $t$ hypersurface of the new coordinates. A similar argument can be followed for the intrinsic curvature $\R_{ij}$. 

A useful summary of the transformation properties of the quantities appearing in unitary gauge is
\begin{align}
f &\to f + \dot f \pi  + \frac12 \ddot f \pi^2 \;,  \label{trans_ST_6} \\
g^{00} &\to g^{00} +  2 g^{0 \mu} \dot \pi + g^{\mu \nu} \partial_\mu \pi \partial_\nu \pi \;, \label{trans_g00} \\
\delta K_{ij} &\to \delta K_{ij} - \dot H \pi h_{ij} - \partial_i \partial_j \pi  \;, \label{transK} \\
\delta K &\to \delta K  - 3 \dot H \pi - \frac1{a^2} \partial^2 \pi \;, \\ 
{}^{(3)}\!R_{ij} &\to {}^{(3)}\!R_{ij}  + H (\partial_i \partial_j \pi + \delta_{ij} \partial^2 \pi) \;, \label{transR}\\
{}^{(3)}\!R &\to {}^{(3)}\!R + \frac4{a^2} H \partial^2 \pi \;. \label{variationR}
\end{align}

Coherently with point {\bf (V)} of Sec.~\ref{Sec_2.1} we note that a way to \emph{not} produce any Goldstone field $\pi$ from actions~\eqref{example2} and~\eqref{example3} is $f = 1$, $\Lambda = const.$, $c = 0$, $S^{(2)} = 0$. In other words, the requirement that time translations are unbroken, in the case of inflation, forces towards the (strict) de Sitter limit. In the case of dark energy the same requirement produces a simple cosmological constant term.

\section{Top-down construction}
\label{sec_4}

 So far we have kept a strict ``bottom-up" perspective that has shown that the proposed action for cosmological perturbations is the natural consequence of the spontaneous breaking of time translations of any cosmological background. Now it is worth giving also a different, perhaps more mundane, perspective by starting from a covariant action for a scalar-tensor theory with fields $\phi$ and $g_{\mu \nu}$ and look at the same action from ``top-down".  
 
\subsection{Simple examples}
 
In a general (perturbed) FRW universe, $\phi(t, \vec x) = \phi_0(t) + \delta \phi(t, \vec x)$. By choosing the coordinate $t$ to be a function of $\phi$, $t = t(\phi)$, we thus simply have $\delta \phi = 0$. Therefore, the action written in this gauge only displays metric degrees of freedom. For instance, 
a canonical kinetic scalar term $(\partial \phi)^2$ is written in unitary gauge as
\begin{equation}\label{g00}
-\frac{1}{2} (\partial \phi)^2 \ \equiv \ - \frac{1}{2} g^{\mu \nu} \partial_\mu \phi \partial_\nu \phi \ \rightarrow \ - c_0(t) g^{00}\, .
\end{equation}
Note however that  $c_0$ is only one of the potentially many contributions to the term $c(t)$ in the actions~\eqref{example2} and~\eqref{example3}. For example, the covariant operator $(\partial \phi)^2 R$ that represents a higher-derivative coupling between the metric and the scalar field can be expanded in perturbations as
\begin{equation}
(\partial \phi)^2 R\ \ = \ \dot \phi_0^2 \left[-R + R^{(0)}(t) + R^{(0)}(t) g^{00} + \delta g^{00} \delta R\right],
\end{equation}
with $R^{(0)}$ the background value of the Ricci scalar. The first three terms in brackets contribute to the EFT terms displayed in~\eqref{example3}, while the forth is already explicitly second order in the perturbations.

By generalizing~\eqref{g00}, it is immediate to see how action \eqref{total_action} includes also $k-$inflation and $k$-essence models \cite{ArmendarizPicon:1999rj,ArmendarizPicon:2000dh}. There, the Lagrangian has at most one derivative acting on each field $\phi$, ${\cal L} = P(\phi, X)$, where $X = \partial_\mu \phi \partial^\mu \phi$ (note that $X$ is sometimes defined with a $-1/2$ factor). In unitary gauge this is of the form $P( \phi_0(t) , \dot \phi_0^2 g^{00})$, which can be expanded in powers of $\dot \phi_0^2  \delta g^{00}$. By redefining the field in such a way that $\phi_0 = t$, it is straightforward to see the various contributions to action  \eqref{total_action},
\be
\Lambda(t)  = c  (t) -P(t, -1) \;, \qquad
c(t)  =  - \left. \frac{\partial P}{\partial X} \right|_{X=-1}\; ,  \qquad
M_n^4 (t)  =  \left. \frac{\partial^n P}{\partial X^n} \right|_{X=-1}  \quad (n\ge 2) \;.
\ee

 In a way, Brans-Dicke~\cite{BD} and $F(R)$ theories~\cite{DeFelice:2010aj,Chiba:2003ir} are even easier to include in this formalism because, at least in their basic versions, they do not need any higher-order operator and are completely described by the operators explicitly displayed in~\eqref{example3}. 

An detailed dictionary for writing covariant operators of increasing complexity in unitary gauge can be found in Sec. 3 of Ref.~\cite{Gleyzes:2013ooa}. In the following subsection we summarize the results by considering the full Horndenski Lagrangian.

\subsection{Horndeski theory}

\newcommand{\Gtwo}{G_2{}}
\newcommand{\Gthree}{G_3{}}
\newcommand{\Gfour}{G_4{}}
\newcommand{\Gfive}{G_5{}}
\newcommand{\Ftwo}{F_2{}}
\newcommand{\Fthree}{F_3{}}
\newcommand{\Ffour}{F_4{}}
\newcommand{\Ffive}{F_5{}}

In four dimensions, the most general scalar-tensor theory having field equations of second order in derivatives is a combination of the generalized Galileon Lagrangians~\cite{horndeski,Deffayet:2009mn,DGSZ},
\be
L = L_2 + L_3 + L_4+ L_5\;, 
\ee
where
\begin{align}
L_2 & =  \Gtwo \;,  \label{Lo2} \\
L_3 & = \Gthree \, \Box \phi \;, \label{L3} \\
L_4 & = \Gfour \,  R - 2 \Gfour_{X}\, (\Box \phi^2 - \phi^{; \mu \nu} \phi_{; \mu \nu}) \;, \label{L4} \\
L_5 & =  \Gfive\,   G_{\mu \nu} \phi^{;\mu \nu} +\frac13  \Gfive_{X} \,  (\Box \phi^3 - 3 \, \Box \phi \, \phi_{;\mu \nu}\phi^{;\mu \nu} + 2 \, \phi_{;\mu \nu}  \phi^{;\mu \sigma} \phi^{; \nu}_{\ ; \sigma}) \;, \label{L5}
\end{align}
and $\Gtwo$, $\Gthree$, $\Gfour$ and $\Gfive$ are functions of $\phi$  and  $X$.

It is possible to translate this theory in the EFT language by first rewriting the above Lagrangian in terms of 3-d geometrical objects induced on uniform $\phi$ hypersurfaces. In particular, we can first define the future directed unitary vector orthogonal to these hypersurfaces. Up to a factor $\gamma$, it  is proportional to the gradient of $\phi$, 
\be
n_\mu \equiv - \gamma \, \phi_{; \mu}, \qquad \gamma \equiv  {1}/{\sqrt{-X}}\, .
\ee
The metric induced on the $\phi= const.$ hypersurface is $h_{\mu \nu} \equiv n_\mu n_\nu + g_{\mu \nu}$. 
Finally, we can define the extrinsic curvature  as $K_{\mu \nu} \equiv h_\mu^\sigma\,  n_{\nu;\sigma} $ and the 3-Ricci tensor computed from the induced metric $h_{\mu \nu}$ as $\R_{\mu \nu}$.
The key ingredient is then to  decompose the covariant derivative of $n_\nu$ as $n_{\nu ; \mu} = K_{\mu \nu} - n_\mu \nd_\nu $, where the acceleration vector $\nd_\mu$ is defined as $\nd_\mu = n^\nu \, n_{\mu ; \nu} $. By means of the quantities just defined, we can finally decompose the second derivative of the scalar field as 
\begin{equation}
\phi_{; \mu \nu} =- \gamma^{-1}(K_{\mu \nu} - n_\mu \nd_\nu - n_\nu \nd_\mu)+ \frac{\gamma^2}{2} \phi ^{; \lambda} X_{; \lambda} n_\mu n_\nu \label{phimunu}\, .
\end{equation}

Making use of this decomposition and of the Gauss-Codacci relation \eqref{GC_equ} and its contractions, after several manipulations it is very lengthy but straightforward to show that the above Lagrangian can be rewritten, up to boundary terms, as \cite{Gleyzes:2013ooa}
\be
\begin{split}
\label{Gali_Lag_EFT}
L &= -\frac{1}{3} (-X)^{3/2} \Gfive_X (K^3 - 3 K K_{\mu \nu}K^{\mu \nu} + 2  K_{\mu \nu}  K^{\mu \sigma} K^\nu_{\ \sigma}) - \sqrt{-X}  \Ffive   \left( K^{\mu \nu} \R_{\mu \nu} - \frac12 K \R \right) \\
 & + \left( 2 X \tilde \Gfour_X - \tilde \Gfour \right)(K^2 - K_{\mu \nu}K^{\mu \nu})  + \tilde \Gfour \R -  \sqrt{-X} (2\Gfour_\phi + 2X \Fthree_X) K - X \Fthree_\phi +\Gtwo \;.
\end{split}
\ee
The auxiliary functions $\Ffive$ and $\Fthree$ are defined  by
\be
\Gthree \equiv  \Fthree + 2 X \Fthree_{X}  \;, \qquad \Gfive_{X} \equiv   \Ffive_{X} + {\Ffive}/({2X})\;, \label{F5_F3}
\ee
and the function 
$\tilde \Gfour \equiv \Gfour + X(\Gfive_\phi - \Ffive_\phi) /2 $
has been introduced to simplify the notation.

In unitary gauge $\phi (t,\vec x)= \phi_0(t)$  the functions $G_i$ and $F_i$ on $(\phi, X)$ become dependent on $(\phi_0(t),  \dot \phi_0^2(t)g^{00})$. These functions can be thus expanded in powers of $\delta g^{00}$ with time-dependent coefficients. It is now straightforward to write the Lagrangian above in unitary gauge in the EFT language by integrating by parts the term linear in $K$ and expanding $K$ and $K_{\mu \nu}$ in the other terms around their background values. One obtains 
\be
\begin{split}
\label{action_Gali_EFT}
S & =S_0 \,+  \int \! d^4x \sqrt{-g} \bigg\{ \,  \frac{M_2^4 (t)}{2} (\delta g^{00})^2\, -\, \frac{m_3^3(t)}{2} \, \delta K \delta g^{00} \, 
-  m_4^2(t) \, \left(\delta K^2 - \d K^\mu_{ \ \nu} \, \d K^\nu_{ \ \mu} - \frac12  \R \, \delta g^{00} \right)   \\
&+    \frac{m_5(t)}3 \left[ \delta K^3 - 3 \delta K \delta K_{\mu \nu} \delta K^{\mu \nu} + 2 \delta K_{\mu \nu} \delta K^{\mu \sigma} \delta K^\nu_{\ \sigma}  - \delta g^{00} \left( K^{\mu \nu} \R_{\mu \nu} - \frac12 K \R \right) \right] + \ldots  \bigg\} \;,
\end{split}
\ee
where the dots $\ldots$ stand for cubic or higher-order terms containing the same four operators explicitly written in the action times  higher powers of $\delta g^{00}$; for instance, $(\delta g^{00})^3$, $\delta K (\delta g^{00})^2$, etc. The explicit relations between the six time-dependent coefficients $f$, $\Lambda$, $c$, $M_2^4$, $m_3^3$, $m_4^2$ is given in Ref.~\cite{Gleyzes:2013ooa}. Here we just note that the three coefficients $\Lambda$, $c$ and $M_2^4$  are affected by all the four  Galilean Lagrangians $L_i$; $m_3^3$ is not affected by $L_2$ while $f$ and $m_4^2$ are only affected by $L_4$ and $L_5$. Finally, $m_5$ is {\em only} affected by $L_5$. Indeed, in unitary gauge $\delta (\sqrt{-X} \Ffive_X ) = \sqrt{-X} \Gfive_X \delta X$, which can be derived from eq.~\eqref{F5_F3}. Using this relation in eq.~\eqref{Gali_Lag_EFT} and comparing it with the Lagrangian in eq.~\eqref{action_Gali_EFT}, one finds
\be
m_5 (t) = - \dot \phi_0^3(t)\, \Gfive_X \left(\phi_0(t),  \dot \phi_0^2(t)g^{00}\right) \;.
\ee
Since $L_4$ and $L_5$ start differing only by the operator proportional to $m_5$ which is   cubic, at quadratic order in the action $L_4$ and $L_5$ carry the same dynamics.
The first line---i.e.~the action up to second order---is equivalent to the first two lines of action \eqref{total_action}, which for $m_4^2= \tilde m_4^2$ contain the set of quadratic operators that are known not to generate higher derivatives  in the linear equations of motion \cite{Gleyzes:2013ooa,Bloomfield:2013efa}. This implies, remarkably, that the dynamics of linear perturbations can be more general than that of Horndeski while remaining second order.

\section{ADM analysis}
\label{sec_3}

Without exiting the unitary gauge, we now perform a complete dynamical analysis of the various quadratic operators of eq.~\eqref{total_action} in the ADM formalism. After solving for the Hamiltonian and momentum constraints, the purpose of this section is to write a quadratic action for the variable $\zeta$, defined in eq.~\eqref{zetaporca} below. Our analysis shows, among other things, that the operators contained in the first two lines of~\eqref{total_action} do not involve higher (time and space) derivatives for the variable $\zeta$. 
Similar analysis in unitary gauge can be found in \cite{malda,EFT1,Boubekeur:2008kn,Gleyzes:2013ooa}.

\subsection{Universal part of the action}

Let us first consider the universal part of action~\eqref{total_action},
\begin{equation} \label{firstS}
S_0 =  \int d^4x \sqrt{-g} \left[\frac{\MM^2}{2} f(t) R - \Lambda(t) - c(t) g^{00}   \right] \;,
\end{equation}
that contains the only operators which are also zeroth and first order in the perturbations.

We will use the ADM formalism to study this action. The ADM metric is
\be
ds^2=-N^2 dt^2 +{h}_{ij}\left(dx^i + N^i dt\right)\left(dx^j + N^j dt\right) \, ,
\ee
where $h_{ij}$ is the induced spatial metric on constant time hypersurfaces and $N$ and $N^i$ are respectively the lapse and the shift.
We decompose $R$ in \eqref{firstS} using the contracting form of the Guass-Codazzi relation \eqref{GC_equ}, 
\be \label{RR}
R = \R + (K_{\mu \nu} K^{\mu \nu} -  K^2) + 2 \nabla_\nu(n^\nu \nabla_\mu n^\mu-n^\mu \nabla_\mu n^\nu) \;,
\ee
and employ the ADM expression for the extrinsic curvature,
\be
\label{Kij}
K_{ij} = \frac{1}{N} E_{ij}\;, \qquad E_{ij} \equiv \frac12 (\dot{h}_{ij} - \nabla_iN_j-\nabla_jN_i) \;,
\ee
where the covariant derivative $\nabla_i$ are taken with respect to the 3-d spatial metric $h_{ij}$ (note that $K^{0 \mu} =0$), and for the upper time-time component of the metric, $g^{00} = - N^{-2}$. 
Integrating by parts the last term on the RHS of~\eqref{RR}, the action becomes
\be
\label{2S}
S_0 =  \int d^4x \sqrt{h} \bigg\{\frac{\MM^2 f }{2}  \left[ N {}^{(3)}\!R + N^{-1}(E_{ij} E^{ij} - E^2) - 2 ({\dot f}/{f})N^{-1} E  \right] - N \Lambda + N^{-1}c   \bigg\} \;.
\ee
The background equations can be obtained by varying the homogenous action with respect to $N$ and $a$ (using $\sqrt{h} = a^3)$. This yields
\begin{align}
3 \MM^2 (H^2 f +H \dot{f} ) & = \Lambda + c  \label{bckgd1} \;, \\
\MM^2 (2 f \dot{H} - H {\dot{f}} +  {\ddot{f}}) & = - 2 c   \label{bckgd2}  \;.
\end{align}

By varying  action \eqref{2S} with respect to $N^i$ and $N$ we find the momentum and Hamiltonian constraint, respectively
\begin{align}
0 & ={\cal P}_{0i} \equiv \nabla_k\big[-\MM^2fN^{-1}(E_i^k-E\d_i^k) + \MM^2\dot{f}N^{-1}\d_i^k\big] \;,  \label{con1}\\
0 & = {\cal H}_{0} \equiv \MM^2 f \big[ \R - N^{-2} (E_{ij} E^{ij} - E^2 ) + 2 ({\dot f}/{f}) N^{-2} E  \big]  - 2 \Lambda - 2 N^{-2} c  \label{con2} \;. 
\end{align}
We only need the linear solution of these equations---second order terms in $N$ or $N^i$ will multiply the constraints and will thus vanish \cite{malda}. We expand $N \equiv 1+ \delta N$ and  decompose the shift into a scalar and a transverse part, $N^i \equiv \partial_i \psi + N_T^i$, with $\partial_i N^i_T=0$. Since here we are only concerned with scalar perturbations we pose (see \cite{Gleyzes:2013ooa} for a derivation of the quadratic action of tensor modes)
\begin{equation} \label{zetaporca}
{h}_{ij}\ =\ a^2(t) e^{2\zeta}\, \delta_{ij}.
\end{equation}
The following expressions, which are exact in unitary gauge, will be also useful,
\begin{align}
\R_{ij} &= - \partial_i  \partial_j \zeta + \partial_i \zeta \partial_j \zeta - \delta_{ij} \big[ \partial^2 \zeta + (\partial \zeta)^2 \big] \;, \\
E^i_j & = \big(H + \dot \zeta - \partial \zeta \partial \psi \big) \delta^i_j - \partial_i \partial_j \psi  - \frac12 (\nabla_{i} N^{j}_T + \nabla^{i} N_{j}^T)\;.
\label{Eij_zeta}
\end{align} 
Solving the momentum constraint at first order 
gives 
\begin{align}
\d N& = \frac{\dot{\zeta}}{A_0} \label{dN} \;, \qquad 
A_0  \equiv H+\frac{\dot{f}}{2f}  \;,\\
N_T^i&=0\;, \label{2const}
\end{align}
Using this equation and the background equation \eqref{bckgd1}, the Hamiltonian constraint yields
\be
\label{psi}
\partial^2 \psi  =  \frac{1}{A_0} \bigg[ \bigg( \frac34 \frac{\dot f^2}{f^2} + \frac{c}{f^2 \MM^2} \bigg) \frac{\dot \zeta}{A_0} -  \frac{\partial^2 \zeta}{a^2}  \bigg]  \;.
\ee

One can expand the action \eqref{2S} up to second order and replace $\d N$ using eq.~\eqref{dN}. We do not need to use the solution of the Hamiltonian constraint, eq.~\eqref{psi}. Indeed, the shift $N^i$ only appears either as a linear term proportional to $\nabla_i N^i$ or in the combination $ \nabla_i N_j \nabla^j N^i - (\nabla_i N^i)^2$. Because of eq.~\eqref{2const}, both these terms can be integrated out of the action. Thus, we find
\be
\label{4S}
\begin{split} 
S_0 = & \ \int d^4x a^3 e^{3\zeta} \bigg\{   \big(1 + \frac{\dot \zeta}{A_0} \big)  \left[  - {\MM^2 f} \big( 2 \partial^2 \zeta + (\partial \zeta)^2 \big) a^{-2} e^{-2\zeta} - \Lambda  \right]   \\
& + \ \frac{1}{\big(1 + \frac{\dot \zeta}{A_0} \big)} \big[ -3\MM^2f (H+\dot{\zeta})^2 -3\MM^2 \dot f (H + \dot \zeta) +c \big] \bigg\} \;,
\end{split}
\ee
where we have used $\sqrt{h} = a^3e^{3\zeta} $.
Collecting all the terms in powers of ${\dot{\zeta}}/{A_0}$, one can use the background equation \eqref{bckgd1} to simplify  terms proportional to $ (\dot \zeta/A_0)^2$ and to show that those proportional to ${\dot{\zeta}}/{A_0}$ vanish. Terms proportional to $ (\dot \zeta/A_0)^0$ also vanish, as one can check using the background equations \eqref{bckgd1} and \eqref{bckgd2} and an integration by parts.
Thus, using again the background equations and the definition of $A_0$ the final action reads
\be
\label{S0}
S_0 =  \int d^4x a^3\left[ \alpha_0 \dot{\zeta}^2 - \beta_0 \frac{1}{a^{2}}(\partial \zeta)^2    \right]\;, \qquad  \alpha_0 \equiv \beta_0 \equiv \frac{1}{A_0^2}\left(c  +\frac34 \frac{\dot{f}}{f}^2\right) \;. 
\ee
As expected, this corresponds to a propagating d.o.f.  with unity sound speed, $c_s^2 \equiv \beta_0/\alpha_0=1$.

\subsection{Quadratic operators}

We can now add all quadratic operators that are known not to generate higher derivatives \cite{Gleyzes:2013ooa} in the linear equations of motion, 
\be
\begin{split}
\label{quad_action}
S =S_0 \,+ & \int \! d^4x \sqrt{-g} \left[\,  \frac{M_2^4}{2} (\delta g^{00})^2\, -\, \frac{m_3^3}{2} \, \delta K \delta g^{00} \, 
-  m_4^2\left(\delta K^2 - \d K^\mu_{ \ \nu} \, \d K^\nu_{ \ \mu} \right) \, +\, \frac{\tilde m_4^2}{2} \, \R \, \delta g^{00}  \right] \;.
\end{split}
\ee
Another operator that does {\em not} generate higher derivatives in the equations of motion is
\be
\R_{\mu \nu} \delta K^{\mu \nu}  - \frac12  \R \, \delta K  \;.
\ee 
However, we did not explicitly include it in eq.~\eqref{quad_action} because {\em at quadratic order} it can be re-expressed as the operator $\tilde m_4^2$ using the relation
\be
\lambda(t) \left(\R_{\mu \nu} \delta K^{\mu \nu}- \frac{1}{2} \R\; \delta K \right)  =\frac{\dot{\lambda}(t)}{4}  \, \R \, \delta g^{00} \;,
\ee
which is valid up to boundary terms (see App. A of Ref.~\cite{Gleyzes:2013ooa}).

Using $\delta K_{ij} = - \delta N H h_{ij} + \delta E_{ij}$ and $\delta g^{00} = 2 \delta N$, variation of the full action with respect to  $N^i$ and $N$ yields the momentum and Hamiltonian constraints,
\begin{align}
 0 & = {\cal P}_{0i} + \nabla_k \big[ 2 m_4^2 (\delta E_i^k - \delta E \delta_i^k) + ( m_3^3 - 4 H m_4^2)\delta N \delta_i^k   \big]  \;, \\
 0 & = {\cal H}_0 + 2 \big( 2 M_2^4 + 9 H m_3^3 - 6 m_4^2 H^2  \big) \delta N - \big( m_3^3 + 4 H m_4^2 \big) \delta E + \tilde m_4^2 \, \R \;,
\end{align}
with ${\cal P}_{0i}$ and ${\cal H}_{0}$ defined in eqs.~\eqref{con1} and \eqref{con2}. Their solutions are
\begin{align}
\delta N  & = \frac{\dot \zeta}{A} \;, \qquad A \equiv  H  + \frac{\MM^2 \dot f - m_3^3}{2 \left(f  \MM^2+ 2 m_4^2\right)} \label{dN_quad}\;, \\
\partial^2 \psi & =  \frac{1}{A} \bigg[ \bigg( \frac32 (A-H)^2   + \frac{c + 4 M_2^4}{f^2 \MM^2+2 m_4^2 } \bigg) \frac{\dot \zeta}{A} - \bigg( \frac{\MM^2 f + 2 \tilde m_4^2}{\MM^2 f + 2  m_4^2} \bigg) \frac{\partial^2 \zeta}{a^2}  \bigg] \;,
\end{align}
and eq.~\eqref{2const}. 

Action \eqref{quad_action} can be expanded up to second order and one can replace $\d N$ using eq.~\eqref{dN_quad}. As in the previous subsection, we do not need to use the solution of the Hamiltonian constraint as the shift $N^i$ only contributes to boundary terms. 
As before, by using the background equations one can check that mass terms cancel, as expected; moreover, terms of the type $\dot \zeta \partial^2 \zeta$ can also be reduced to the form $(\partial \zeta )^2$ after  integrations by parts. We finally obtain
\be
S =  \int d^4x a^3\left[ \alpha \dot{\zeta}^2  - \beta \frac{1}{a^2}(\partial \zeta)^2 \right]\;, 
\ee
with 
 \begin{align}
 \alpha& \equiv \frac1{A^2} \left[c + 2 M_2^4 + \frac34 \frac{(\MM^2 \dot f  - m_3^3)^2}{ \MM^2 f+ 2 m_4^2} \right] \;, \label{alpha}\\
 \beta& \equiv - \MM^2 f  + \frac{1}{2 a} \frac{d}{dt} \left[\frac{2 (\MM^2 f + 2 \tilde m_4^2 )a}{A}  \right]\;. \label{beta}
 \end{align}
Stability (absence of ghosts) is ensured by the positivity of  $\alpha$, eq.~\eqref{alpha}, i.e.~the coefficient in front of the time kinetic term. 
The speed of sound squared is given by $c_s^2 = \beta/\alpha$ and its expression simplifies by use of the background equation of motion \eqref{bckgd2} when $m_4^2 = 0 = \tilde m_4^2$, in which case \cite{GPV}
\be \label{cies}
c_s^2 = \frac{c + \frac34  \MM^2 {\dot f^2}/{f} - \frac12 m_3^3 {\dot f}/{f} - \frac14 { m_3^6}/( {\MM^2} f)  + \frac12\left( \dot{ m}^3_3 + H  m_3^3\right) }{c + 2 M_2^4 + \frac34  \MM^2 {\dot f^2}/{f} - \frac32  m_3^3 {\dot f}/{f} + \frac34 { m_3^6}/ ( {\MM^2} f) }\;.
\ee

\subsection{Higher spatial derivatives}

As mentioned earlier, the quadratic operators appearing in action \eqref{quad_action} do not yield higher derivatives in the linear dispersion relation. In particular, it is straightforward to verify using eqs.~\eqref{Kij} and \eqref{Eij_zeta} that $ \delta K^2$ contains a higher spatial derivative term,  $(\partial^2 \psi)^2$, while $ \delta K^\mu_{\ \nu} \d K_\mu^{\ \nu}$  contains $(\partial_i \partial_j \psi)^2$. However, taken in the combination as in eq.~\eqref{quad_action}, these higher-derivative terms combine and give an irrelevant boundary term. 

Independent operators that generate higher spatial---but not time---derivatives in the linear equations of motion are
\be
\label{hsd}
S_{\rm h.s.d.} = \int \! d^4x \sqrt{-g} \left[ - \,  \bar m_4^2(t)\, \delta K^2 \, +\, \frac{\bar m_5(t)}{2} \, \R \, \delta K \, + \frac{\bar \lambda(t)}{2} \R^2 \right] \; .
\ee
We have already mentioned $ \delta K^2$. The operator $\R \, \delta K$ contains  $\partial^2 \psi \partial^2 \zeta$ and, finally, $\R^2 = 16 (\partial^2 \zeta)^2/a^4 $ so that they are both higher-derivative terms. Note that  $ \R_{\mu \nu} \R^{\mu \nu} = [ 5 (\partial^2 \zeta)^2 + (\partial_i \partial_j \zeta)^2] /a^4 $. Thus, to quadratic order it can be rewritten as $\R^2$ up to a total derivative. 
Finally, one could take quadratic combinations of the 3-d Riemann tensor such as $\R_{\mu \nu \rho \sigma} \R^{\mu \nu \rho \sigma}$. However, in three dimensions the Riemann tensor can be expressed in terms of the Ricci scalar and tensor.\footnote{This can be done using the relation  \be
\R_{\mu \nu \rho \sigma} = \R_{\mu \rho} h_{\nu \sigma} - \R_{\nu \rho} h_{\mu \sigma} - \R_{\mu \sigma} g_{\nu \rho} + \R_{\nu \sigma}h_{\mu \rho} -\frac12 \R (h_{\mu \rho} h_{\nu \sigma} - h_{\mu \sigma} h_{\nu \rho}) \;.
\ee} Thus, at quadratic order in the perturbations, 
actions \eqref{quad_action} and \eqref{hsd} seem to exhaust  all the possible independent operators.

When one of these operators is present in the action the dispersion relation of the propagating mode receives corrections $\propto k^4$ at large momenta, so that the dispertion relation becomes  $\omega^2 = c_s^2 k^2 + k^4/M^2$, where $M$ is a mass scale. These corrections may become important in the limit of vanishing sound speed, such as in the model of the Ghost Condensate~\cite{ghost} or for deformations of this particular limit \cite{EFT1,Creminelli:2008wc}.

\section{Inflation and non-Gaussianities}

The EFT for cosmological perturbations turns out to be  enlightening and  useful for inflation, especially for the computation of primordial non-Gaussianity, i.e.~the 3- or 4-point correlation functions of the curvature perturbation $\zeta$. Without  the pretence of being exhaustive, here we discuss few of the main ingredients intervening in the application of this approach to inflation.

As discussed in Sec.~\ref{sec_2}, in the absence of matter fields one can always get rid of the time-dependent  function $f(t)$ in front of $R$ on the first line of eq.~\eqref{total_action}  by  an appropriate field redefinition $g_{\mu \nu} \to  f(t) g_{\mu \nu}$ \cite{EFT2}. This corresponds to going to the so-called Einstein frame. In this frame, the general quadratic and higher-order action is still given by the second line and below of eq.~\eqref{total_action}, but the coefficients in front of the operators get redefined by this transformation. The explicit redefinition is given in detail in Ref.~\cite{GPV}. Moreover, by combining eqs.~\eqref{c2} and \eqref{L2} with eqs.~\eqref{frie1} and \eqref{frie2} for $f=const.$ and setting $M^2_{\rm Pl} \equiv \MM^2 f$, one obtains \cite{EFT1,EFT2}
\be
\label{cL_inflation}
c(t) =  - M^2_{\rm Pl} \dot H \;, \qquad \Lambda (t) = M^2_{\rm Pl} (3H^2 + \dot H) \;.
\ee
Thus, the inflationary background univocally fixes the functions $c(t)$ and $\Lambda(t)$.

As discussed in Sec.~\ref{sec:gravity},  we can associate a Goldstone boson $\pi$ to spontaneously broken time translations during inflation.
Describing inflationary fluctuations in terms of this field greatly helps obtaining the leading order results for the 2-point and higher-order correlation functions. Indeed, in the limit of high energy the Goldstone boson decouples from gravity. This is analogous to what  happens in the gauge theory with  non-Abelian gauge group $A_\mu$ discussed in Sec.~\ref{Sec_2.1}. In this case one can see from eq.~\eqref{L_A_quadratic} that
in the limit $m \to 0$ and $g \to 0$ keeping $m/g$ constant, the Goldstone bosons decouple from the gauge fields $A_\mu$. In other words, at high energies $E \gg m$ it is convenient to use $\pi$ to describe the scattering of massive vector fields, as implied by the equivalence theorem for the longitudinal components of a massive gauge boson \cite{Cornwall:1974km}. In writing \eqref{L_A} as \eqref{L_A_quadratic} we have neglected cubic and higher-order terms in $\pi_c$ suppressed by $m^2/g^2$, suggesting that the Goldstone boson self-interactions become strongly coupled at energies $E \gg 4 \pi m/g$. The decoupling limit is thus well-defined in the regime $m \ll E \ll 4 \pi m /g$.

Let us go back to inflation and introduce the field $\pi$ via the St\"uckelberg trick. We consider the action \eqref{total_action} but we first neglect operators containing the extrinsic curvature and the 3-Ricci scalar. Using eq.~\eqref{variationR}  for this action one obtains \cite{EFT2}
\be
\begin{split}
S = \int \! d^4 x   \sqrt{- g} &\left[\frac{M_{\rm Pl}^2}{2} R
- M^2_{\rm Pl} \left(3H^2(t+\pi) +\dot{H}(t+\pi)\right)+ \right.\\ 
&+M^2_{\rm
Pl} \dot{H}(t+\pi)\left(
(1+\dot\pi)^2g^{00}+2(1+\dot\pi)\partial_i\pi g^{0i}+
g^{ij}\partial_i\pi\partial_j\pi\right) + \\ \nonumber
&\frac{M_2^4(t+\pi)}{2}\left(
(1+\dot\pi)^2g^{00}+2(1+\dot\pi)\partial_i\pi g^{0i}+
g^{ij}\partial_i\pi\partial_j\pi+1\right)^2 + \nonumber\\
\nonumber & \left. \frac{M_3^4(t+\pi)}{3!}\left(
(1+\dot\pi)^2g^{00}+2(1+\dot\pi)\partial_i\pi g^{0i}+
g^{ij}\partial_i\pi\partial_j\pi+1\right)^3+ ... \right] \; .
\end{split}
\ee
The first two lines of this action can describe a canonical scalar field rolling along its potential, i.e.~slow-roll inflation. The leading mixing with gravity comes from the operator
\be
M_{\rm Pl}^2 \dot H \dot \pi \delta g^{00} \sim \dot H^{1/2} \dot \pi_c \delta g^{00}_c \;,
\ee
where for the second approximate equality we have defined the canonically normalized fields $\pi_c \equiv M_{\rm Pl} \dot H^{1/2} \pi$ and $\delta g^{00}_c \equiv M_{\rm Pl} \delta g^{00}$. In analogy with the gauge theory case discussed above, the decoupling limit thus corresponds to the limit where the coupling constant and the mass go to zero, respectively  $g = M_{\rm Pl}^{-1} \to 0$ and $m = \dot H^{1/2} \to 0$, while keeping $m/g = M_{\rm Pl} \dot H^{1/2}$ constant.

If $M_2^4 $, $M_3^4 $, etc.~do not vanish, the action above can describe a derivative expansion of the inflaton field or, more generally, a Lagrangian which depends both on $\phi$ and on $X$, such as $k$-inflation \cite{ArmendarizPicon:1999rj}. Such a non slow-roll dynamics takes place when $M_2^4 \gg M_{\rm Pl}^2 \dot H $, in which case the mixing with gravity is dominated by the operator,
\be
M_2^4 \dot \pi \delta g^{00} \sim ( M_2^2 /M_{\rm Pl}) \dot \pi_c \delta g^{00}_c \;,
\ee
where this time we have defined $ \pi_c \equiv M_2^2 \pi$. The analogy with the gauge theory and the decoupling energy is again straightforward once we identify $g = M_{\rm Pl}^{-1} $ and $m =  M_2^2 /M_{\rm Pl}$. We conclude that at high energy, $E \gg m$, neglecting the mixing with gravity and mass terms of $\pi$, the action of the Goldstone boson simplifies, in the notation of \cite{Senatore:2009gt}, to
\be 
\label{action_pi_infl}
S_{\rm \pi} = \int \!  d^4 x   \sqrt{- g} \frac{(- M^2_{\rm Pl} \dot{H}) }{c_s^2}
 \left[  \dot\pi^2 - c^2_s \frac{ (\partial_i \pi)^2}{a^2}
-(1-c_s^{2}) \dot \pi \frac{ (\partial_i \pi)^2}{a^2}
+ (1-c_s^{2}) \left(1+ \frac23 \frac{\tilde c_3}{ c_s^2} \right) \dot \pi^3 \right] \;,
\ee
where we  define the sound speed of fluctuations $c_s^2$ and the parameter $\tilde c_3$ by
\be
c_s^{-2} \equiv 1- \frac{2 M_2^4}{M_{\rm Pl}^2 \dot H} \;, \qquad \tilde c_3 \equiv - \frac{M_3^4}{M_2^4} c_s^2\;,
\ee
and we have neglected  quartic and higher terms. 

Normalizing the Goldstone boson to the standard Bunch-Davies vacuum on small scales \cite{Birrell:1982ix,Garriga:1999vw}, the negative frequency solution of the wave equation for $\pi$ is, up to slow-roll corrections,
\be
\pi_{\vec k} (\eta) = \frac{c_s}{a M_{\rm Pl} |\dot{H} |^{1/2}}\frac{e^{- i c_s k \eta}}{\sqrt{2 c^3_s  k^3}} \,  (1+ i c_s k \eta) \;,
\ee
where $\eta$ is the conformal time, $\eta \equiv \int dt/a(t)$. Using $\zeta = -H \pi$, on super horizon scales, i.e.~for $- c_sk \eta \ll 1$, the power spectrum of the curvature perturbation reads\footnote{The role of symmetry breaking scale here is played by the combination $(M^2_{\rm Pl} \dot{H} c_s)^{1/4}$, so that the amplitude of the 2-point function of $\zeta$ is set by the ratio between the size of the quantum fluctuations to the symmetry breaking energy to the fourth power. For an exhaustive discussion of all the scales that are relevant during inflation we refer the reader to~\cite{daniel}.}
\be
\langle \zeta_{\vec k} (\eta) \zeta_{\vec k'} (\eta) \rangle = (2 \pi)^3 \delta( \vec k + \vec k') \, \frac{1}{2 k^3} \left.\frac{H^4}{M^2_{\rm Pl}| \dot{H} |c_s} \right|_{-c_s k \eta =1}\;.
\ee
The 3-point function can be computed using the standard machinery of primordial non-Gaussianity (see for instance \cite{malda,Chen:2006nt}). Its amplitude, which is typically given in terms of the nonlinear parameter $f_{\rm NL}$, can be simply estimated by comparing the cubic to the quadratic part of the Lagrangian \cite{EFT2,Leblond:2008gg}. For instance, for the first cubic term in eq.~\eqref{action_pi_infl} one has
\be
\frac{{\cal L}_{\dot \pi (\partial_i \pi)^2 }}{{\cal L}_2}  \sim - \frac{(1-c_s^2) \dot \pi (\partial_i \pi)^2 }{\dot \pi^2} \sim -\frac{1-c_s^2}{c_s^2} \zeta\;, 
\ee
where $\zeta$ denotes the amplitude of the curvature power spectrum, $\zeta  \sim H^2 / (2 M_{\rm Pl}^2 |\dot H| c_s)^{1/2}$, while for the second term one finds ${\cal L}_{\dot \pi^3 }/{\cal L}_2 = - {{\cal L}_{\dot \pi (\partial_i \pi)^2 }}/{{\cal L}_2} (c_s^2 + 2 \tilde c_3 /3)$.
Thus, as $f_{\rm NL} \sim \frac{{\cal L}_3}{{\cal L}_2} \zeta^{-1}$  one has large non-Gaussianity in the limit of small sound speed.

The amplitude of non-Gaussianity is related to the  energy scale at which the theory becomes strongly coupled. This is given by $ \Lambda^4_{\dot \pi (\partial_i \pi)^2} = 16 \pi^2 M_{\rm Pl}^2 |\dot H| c_s^5 (1-c_s^2)^{-2}$ and $ \Lambda^4_{\dot \pi^3} = \Lambda^4_{\dot \pi (\partial_i \pi)^2} (c_s^2 + 2 \tilde c_3 /3)^2$ \cite{Senatore:2009gt,Baumann:2011dt,daniel},\footnote{To estimate the strong coupling scale we can rewrite action \eqref{action_pi_infl} in terms of $\pi_c \equiv M_{\rm Pl} |\dot H |^{1/2} c_s^{-1} \pi $ as
\be
S_{\rm \pi} = \int \!  d^3 x dt   
 \left[  \dot\pi_c^2 - c^2_s { (\partial_i \pi_c)^2} 
-\frac{1}{M^2_{\dot \pi (\partial_i \pi)^2}} \dot \pi_c { (\partial_i \pi_c)^2}
+ \frac{1}{M^2_{\dot \pi^3 }}  \dot \pi_c^3 \right] \;,
\ee
where for simplicity we have neglected the  expansion of the universe and defined $M^2_{\dot \pi (\partial_i \pi)^2} \equiv M_{\rm Pl} |\dot H |^{1/2} c_s^{-1} (1-c_s^2)^{-1}$ and $M^2_{\dot \pi^3} \equiv M^2_{\dot \pi (\partial_i \pi)^2 } (1+\frac23 \frac{\tilde c_3}{c_s^2})^{-1}$. Given the dispersion relation, $\omega = c_s k$, one can check that $\pi_c$ scales as $\propto c_s^{-3/2} \omega$, so that cubic terms become as important as the quadratic ones respectively when $\omega^ 4 \sim  c_s^7 M^2_{\dot \pi (\partial_i \pi)^4}$ and $ \omega^ 4 \sim  c_s^7 M^2_{\dot \pi^3}$. The extra  factor $(4 \pi)^2$ can be found by more formally defining the strong coupling scale as the breakdown of perturbative unitarity of the scattering of $\pi$.
See for instance Appendix E of \cite{daniel}.}  
so that
\be
f_{\rm NL} \sim 8 \pi \left( \frac{H}{\Lambda_3} \right)^2 \zeta^{-1}\;.
\ee
Large non-Gaussianity thus means  that the energy scale of inflation, $H$, is getting close to the strong-coupling scale of the theory~\cite{Leblond:2008gg}. Indeed, constraints on the two parameters space $(c_s^2,\tilde c_3)$ have been put by the Planck satellite mission \cite{Ade:2013ydc}. A well-studied example where this limit is protected by radiative corrections is the UV complete DBI inflation \cite{Silverstein:2003hf,Alishahiha:2004eh}, where $\tilde c_3 = 3 (1-c_s^2) /2. $\footnote{It has been shown that this relation also more simply follows from the $ISO(4,1)$ symmetry in the EFT of inflation \cite{Creminelli:2013xfa}}

\section{Dark energy and modified gravity}

As opposed to inflation, the present acceleration of the universe~\cite{costas,lucashin} does not need an ``ending" mechanism and, as far as we know, could as well last forever.  This makes the case for a scalar degree of freedom more circumstantial. However, every concrete alternative to the cosmological constant involves, in a way or another, a new scalar degree of freedom. 
Even the strongest prejudice towards a simple---though ridiculously fine-tuned---vacuum energy should not prevent an efficient parameterization of general models of dark energy, if nothing else because we need to ``quantify" its observational evidence among all possible alternatives. Various interesting parameterization of the dark energy behavior that are not directly related to the present formalism can be found,  for instance, in Refs.~\cite{Crittenden:2007yy,post-fried2,PZW,Baker:2011jy,BF,Jimenez:2012jg,Battye:2012eu,Baker:2012zs,Sawicki:2012re,Silvestri:2013ne}.

One of the main advantages of the EFT approach~\cite{GPV,BFPW,Gleyzes:2013ooa,Bloomfield:2013efa} is that of offering a clear separation between the background quantities (essentially, the scale factor $a(t)$ as a function of the time) and the effects that dark energy can induce at the level of the perturbations. 
In this formalism these two aspects, that typically correspond to very different observables and experiments, are naturally separated because they are related to different operators.
In particular, the background evolution depends only on the three functions of the time $f(t)$, $c(t)$ and $\Lambda(t)$ through equations~\eqref{c2}-\eqref{frie2}.  However, as opposed to inflation where $H$ and $\dot H$ completely determine the two parameters $c$ and $\Lambda$, here we have to deal with one more degree of freedom, represented by the function $f(t)$. A non-constant function $f$, together with the coefficients of certain quadratic operators, can be responsible for departures from General Relativity, as we show in the following.

As shown by the ADM analysis in Sec.~\ref{sec_3}, theoretical constraints---the number of physical degrees of freedom and their classical and quantum stability---and the linear  dispersion relation of perturbations are directly dictated by the three operators above and the quadratic ones. For instance, for certain combinations of mass coefficients, $\alpha$ and $\beta$ in eqs.~\eqref{alpha} and \eqref{beta} are such that $\beta\ll \alpha$, in which case the sound speed of dark energy becomes very small. The most well-known example is  the case  $c \ll M_2^4$ for $k$-essence or the ghost condensate theory \cite{ghost} and small deviations from its limit \cite{Creminelli:2008wc}.
Interestingly, as long as the scalar field description remains valid, the EFT of perturbations for $c_s^2 \to 0$ applies also in the non-linear regime, i.e.~when the dark energy density becomes non-linear, in which case it can lead to very distinguishable signatures \cite{ArkaniHamed:2005gu,Creminelli:2009mu,Lim:2010yk,Sefusatti:2011cm,D'Amico:2011pf}.

As another simple application, it is worth mentioning the case of a violation of the null energy condition or, in other words, of an effective equation of state for dark energy $w<-1$. A minimally coupled  scalar field with canonic kinetic term cannot reproduce such a situation, if not by brutally appearing in the Lagrangian with the ``wrong" sign for kinetic term and thus immediately leading to ghost excitations. It was soon realized that a sensible theory is possible in the presence of a non-minimal coupling to gravity of the Brans-Dicke type~\cite{Das:2005yj}. Other couplings were also considered, for instance in~\cite{EFT1,Creminelli:2008wc}. Within the present formalism this question basically reduces to an algebraic problem. One has to require an effective super-acceleration at the level of the background equations~\eqref{c2},~\eqref{frie1} and~\eqref{frie2} (for instance: $\dot f = 0$, $c<0$) and then require the time kinetic Lagrangian for the fluctuations to have the good sign by the addiction of appropriate quadratic operators. In practice, the coefficient $\alpha$ of eq.~\eqref{alpha} must be positive.

\subsection{Mixing with gravity}

No obvious distinction between modifying gravity and simple quintessence can be made at the level of the ``unifying" action~\eqref{total_action}. Having decided to write everything in unitary gauge, action~\eqref{total_action} is just the most general option: a generic functional of the metric in the presence of broken time translations and compatible with the residual unbroken three-dimensional space-diffeomorphisms. Whether or not the operators in the action~\eqref{total_action} display departures from General Relativity is ultimately encoded in the behaviour of the probes---the matter fields---under the influence of the metric $g_{\mu \nu}$. A more direct way of studying departures from General Relativity is that of making explicit the scalar degree of freedom of the theory as we did in the last section for inflation, and see what type of coupling it has with the metric field. If this coupling is at the level of the kinetic terms, this is a smoking gun for genuine modifications of gravity. 

As we did for inflation in the last section, in order to make the scalar degree of freedom explicit we apply  the St\"uckelberg trick, i.e.~we force a diffeomorphism $t \rightarrow t + \pi(x)$ upon the unitary gauge action~\eqref{total_action}, as outlined in Sec.~\ref{stuck}. 
The simplest way to generate a dynamical $\pi$ field is to consider a non-vanishing coefficient $c(t)$. In this case the St\"uckelberg trick generates $\pi$ with a relativistic kinetic Lagrangian $\dot \pi^2 - (\vec \nabla \pi)^2$.  A more involved example  is constituted by  the operator $M_2^4$. In order to fix the ideas, once we have moved out of unitary gauge through St\"uckelberg, let us consider  scalar linear perturbations in Newtonian gauge, which is frequently used for late time cosmology,
\begin{equation} \label{newtonian} 
ds^2 = -(1+2\Phi)dt^2 + a^2(t) (1-2 \Psi) \delta_{i j} dx^i dx^j\, .
\end{equation}
By making use of eq.~\eqref{trans_g00} and of the expression for $g^{00}$, one finds $\delta g^{00} \to 2(\Phi -\dot \pi) + 4 \Phi \dot \pi - \dot \pi^2 + a^{-2}(\vec  \nabla \pi)^2$. Thus, the Lagrangian of $\pi$ reads
\be
-c \; \delta g^{00} + \frac{M_2^4}{2} (\delta g^{00})^2 \ = \  (c + 2 M_2^4) \dot \pi^2 -  c ( \vec \nabla \pi)^2   - 4(c+ M_2^4) \dot \pi \Phi + \dots \;.
\ee  
We see that $M_2^4$ does not mix $\pi$ with gravity at the highest energies. The first coupling appears at the level of terms that are quadratic in the fields but with only one derivative
in total. Therefore, at high energy the last term can be neglected, $\pi$ decouples from gravity and propagates with a speed of sound $c_s^2 = c/(c + 2 M_2^4)$. As previously discussed for inflation, this is the so-called decoupling limit \cite{EFT2}, which takes place at an energy higher than $E_{\rm mix} \sim (c + M_2^4)/[(c + 2 M_2^4)^{1/2} \MM]$. 

For other operators,  decoupling is not necessarily at work and $\pi$ and gravity may be mixed already at the kinetic level. 
This can be verified by inspection, considering the explicit (linear) expressions of the curvatures in Newtonian gauge,
\be
\begin{split}
\label{KijRij}
K_{ij} &=  e^{-\Phi} (H-\dot \Psi) h_{ij}   \;, \\
{}^{(3)}\!R_{ij} &= \partial_i \partial_j \Psi + \delta_{ij}  \partial^2 \Psi  \;.
\end{split} 
\ee
So, for instance, the operator $m_3^3$ after the St\"uckelberg trick~\eqref{trans_g00}-\eqref{transK} gives
\begin{equation}
- \frac{m^3_3}{2} \delta K \delta g^{00} \ \rightarrow \ - m^3_3\, \big(3 \dot \Psi \dot \pi + a^{-2} \vec \nabla \pi \vec \nabla \Phi + a^{-2} \dot \pi \nabla^2 \pi  + \dots \, \big)\;,
\end{equation}
where the ellipsis stand for terms of lower order in derivatives. The presence of kinetic mixing between $\pi$ and the gravitational perturbation changes the structure of the theory already at the level of the propagator. The specific type of modification of gravity that the operator $m_3^3$ is responsible for has been named ``kinetic gravity braiding"~\cite{Deffayet:2010qz}, although it was previously studied in \cite{EFT1,EFT2,Creminelli:2008wc}.

A more standard kinetic mixing is the one provided by a non-constant $f(t)$,
\begin{equation}
f (t) R  \ \rightarrow \ 
2 f \left[ - 3 \dot \Psi^2  - 2 \vec \nabla \Phi \vec \nabla \Psi + (\vec \nabla \Psi)^2 + 3 ({\dot f}/{f}) \dot \Psi \dot \pi - ({\dot f}/{f}) \pi (\nabla^2 \Phi - 2 \nabla^2 \Psi)\right]\, .
\end{equation}
This is nothing else than a modification of gravity of the Brans-Dicke type~\cite{BD}.

\subsection{Observables in the perturbation sector}

The above are just two examples of the universality and generality of the EFT approach. Just in terms of few operators one has all the relevant effects that have been studied at length by specific explicit models. All versions and types of modifications of gravity are distilled in a finite number of terms. One can then wonder what are, more in detail, the cosmological consequences of the various operators. By briefly reviewing the more general results of~\cite{Gleyzes:2013ooa} (see, e.g.~also \cite{DeFelice:2011hq,BFPW,Silvestri:2013ne}), here we limit ourselves to the operators of the second line of~\eqref{total_action}: those that do not give higher derivatives in the equations of motion. 

An ambitious target of the future galaxy surveys such as EUCLID~\cite{euclid1,euclid2} and BigBoss~\cite{bigboss} is that of constraining the linear growth factor that determines the growth rate of the large scale structures.  On these scales, for models with $c_s \sim {\cal O} (1)$ we can take the quasi-static approximation, i.e.~neglect anisotropic stresses and the time derivatives in the equations of motion. In this case the evolution of perturbations is described by
\begin{equation}
\label{qs}
{\cal M}_{ab} \, V_b
 = \delta_{a3} \, 
\bar \rho_m \, \Delta_m
\;,
\end{equation}
where $V^a = (\Phi, \Psi, \pi)$, $\bar \rho_m$ and $\Delta_m$ are respectively  the unperturbed density and the density contrast in comoving gauge of non-relativistic matter  and ${\cal M}$ is a matrix given in terms of the time-dependent coefficients in front of the quadratic operators (see \cite{Gleyzes:2013ooa} for details).
The effects of modification of gravity on the linear growth factor  are encoded in a Poisson equation with a modified Newton constant,
\be
-\frac{k^2}{a^2} \Phi \equiv 4 \pi G_{\rm eff}(t,k) \bar \rho_m \Delta_m \;,
\ee
where, using eq.~\eqref{qs}, $G_{\rm eff}(t,k)$ can be written once and for all in terms of the quadratic operators, $4 \pi G_{\rm eff} = - [ {\cal M}^{-1}]_{13} $.
If no higher-derivative terms are considered, it has the following structure,
\begin{equation}
G_{\rm eff}(t,k) \ =\ G_{\rm eff}^{(0)}(t) \ + \  G_{\rm eff}^{(-2)}(t) \left(\frac{k}{a}\right)^{-2} + \, \dots \;,
\end{equation}
for $k \gg a \sqrt{G_{\rm eff}^{(0)}/G_{\rm eff}^{(-2)}}$.
Already the renormalization of the Newton constant  at high momenta---i.e.~$G_{\rm eff}^{(0)}(t)$---signals the we are in the presence of a modification of gravity. 
Another quantity often used to parameterize deviations from General Relativity  is the  ratio between the gravitational potentials $\gamma\equiv \Psi/\Phi$ which, using again
eq.~\eqref{qs}, is given by
$\gamma = {[{\rm com}( {\cal M})]_{32}}/{[{\rm com} ({\cal M})]_{31}} $, where ${\rm com} ( {\cal M})$ denotes the comatrix of ${\cal M}$.
Schematically, at the lowest order in derivatives we have 
\begin{equation}
\gamma \ =\ \gamma^{(0)} \ + \  \gamma^{(-2)} \left(\frac{k}{a}\right)^{-2} + \, \dots \;,
\end{equation}
for $k \gg a \sqrt{\gamma^{(0)}/\gamma^{(-2)}}$. The actual coefficients in terms of the various operators are rather complicated and it is not worth reproducing them here. They can be found in~\cite{Gleyzes:2013ooa} where indeed also the contribution of higher-derivative terms were considered. 

\section{Concluding remarks}
Inflation and dark energy are two of the most challenging aspects of the current picture of the Universe. They are the main target of future high-precision cosmological observations and the subject of a frenetic theoretical activity. In this paper we have reviewed a powerful formalism for cosmological perturbations in the presence of a scalar degree of freedom and outlined its applications for the study of both such epochs of accelerating expansion. Here is a summary of its main features. 
\begin{itemize}
\item The EFT of cosmological perturbations displays a universal action~\eqref{total_action} already expanded in number of perturbations and with no field-redefinition ambiguities. 
\item The action is built in unitary gauge (Sec.~\ref{sec_2}), in which the scalar degree of freedom is ``eaten" by the metric and the expansion in number of perturbations around a FRW background is particularly natural.  
\item Use of the St\"uckelberg trick (Sec.~\ref{stuck}) allows to re-write the action and the equations of motion in any other desired set of coordinates. However, a complete dynamical analysis can be done directly in unitary gauge (Sec.~\ref{sec_3}) by using the ADM formalism.
\item Only three (two, in the case of inflation) time-dependent coefficients determine the background evolution, all other operators have only effects on the dynamics of the perturbations.
\item The theoretical features of  specific models---such as the various kinds of modification of gravity or the non-Gaussian features in the power spectrum---can be traced back to certain specific operators that are quadratic and higher order in the perturbations.  In Sec.~\ref{sec_4} we show how to translate a given scalar field model into the EFT language. 
\item The entire Horndeski theory~\eqref{Lo2}-\eqref{L5}, containing four arbitrary functions of the scalar $\phi$ and its kinetic term $\partial \phi^2$, is described at linear order in this formalism by only six arbitrary functions of the time.  
\end{itemize}

Despite all these promising features and applications of the EFT formalism, a number of issues remain currently open, especially for what concerns applications to dark energy. For instance, it is still unclear how to efficiently incorporate in this approach screening mechanisms \cite{Vainshtein:1972sx,Khoury:2003aq} which can be at work on small scales to evade solar system constraints. Moreover, apart from very simple cases discussed here, the specific observational effects of the various mass scales in front of the operators is still lacking. We expect  these issues to be the object of future studies.

%
\vspace{1cm}

{\bf Acknowledgements:} Exposure to Daniel Baumann, Paolo Creminelli, Alberto Nicolis and Leonardo Senatore is among the causes of our involvement in this subject. We are also very grateful to our collaborators Jerome Gleyzes, Giulia Gubitosi and David Langlois, who helped us understanding many of the things that we have tried to review and summarize here.   F.V.~thanks the PCCP in Paris for kind hospitality during the completion of this work. Moreover, he acknowledges partial support by the ANR {\it Chaire d'excellence} CMBsecond ANR-09-CEXC-004-01. \\

\end{document}